\documentclass{elsart}
\usepackage{latexsym,amsmath,tabularx,amssymb,bm}
\usepackage{epsfig}
\usepackage{citesort}

\long\def\comment#1{ }

\usepackage{graphicx}
\usepackage{amssymb}
\usepackage{makeidx}
\usepackage{graphicx}
\usepackage{multicol}

\newcommand{\nn}{\nonumber\\ }
\newcommand{\beq}{\begin{eqnarray}}
\newcommand{\eeq}{\end{eqnarray}}
\newcommand{\be}{\begin{eqnarray}}
\newcommand{\ee}{\end{eqnarray}}

\newcommand{\tr}{{\rm tr}}
\newcommand{\Tr}{{\rm Tr}}

\newcommand{\cal}{\mathcal} 
\newcommand{\abar}{\bar{\alpha}_s}
\newcommand{\lan}{\langle}
\newcommand{\ran}{\rangle}

\newcommand{\mean}[1]{\left\langle #1 \right\rangle_Y}

\newcommand{\dk}[3]{\frac{(\bm{#1}-\bm{#2})^2}
    {(\bm{#1}-\bm{#3})^2 (\bm{#3}-\bm{#2})^2}}
\newcommand{\del}{\partial}
\newcommand{\atpi}{\frac{\abar}{2\pi}}

\newcommand{\V}{\widetilde V}

\newcommand{\x}{\bm x}
\newcommand{\y}{\bm y}
\newcommand{\uu}{\bm u}

\newcommand{\z}{\bm z}

\def\simge{\mathrel{%
   \rlap{\raise 0.511ex \hbox{$>$}}{\lower 0.511ex \hbox{$\sim$}}}}
\def\simle{\mathrel{
   \rlap{\raise 0.511ex \hbox{$<$}}{\lower 0.511ex \hbox{$\sim$}}}}
\def\bigs{\mathrel{
   \rlap{\raise 0.531ex \hbox{$>$}}{\lower 0.531ex \hbox{$<$}}}}

\def\del{\partial}                              

\begin{document}
\begin{flushright}
~\vspace{-1.25cm}\\
{\small\sf SACLAY--T05/031}
\end{flushright}
\vspace{0.8cm}
\begin{frontmatter}

\parbox[]{16cm}{ \begin{center}
\title{Duality and Pomeron effective theory for QCD at high energy and large $N_c$}

\author{J.-P. Blaizot\thanksref{th2}},
 \address{ECT*, Villa Tambosi, Strada delle Tabarelle 286, I-38050 Villazzano(TN) Italy}
\author{\ E. Iancu\thanksref{th2}},
\author{\ K. Itakura} and
\author{\ D.N. Triantafyllopoulos}

\address{Service de Physique Th\'eorique, CEA/DSM/SPhT,  Unit\'e de recherche
associ\'ee au CNRS (URA D2306), CE Saclay,
        F-91191 Gif-sur-Yvette, France}

\thanks[th2]{Membre du Centre National de la Recherche Scientifique
(CNRS), France.}

\date{\today}
\vspace{0.8cm}
\begin{abstract}
We propose an effective theory which governs Pomeron dynamics in
QCD at high energy, in the leading logarithmic approximation, and
in the limit where $N_c$, the number of colors, is large. In spite
of its remarkably simple structure, this effective theory
generates precisely the evolution equations for scattering
amplitudes that have been recently deduced from a more complete
microscopic analysis. It accounts for the BFKL evolution of the
Pomerons together with their interactions: dissociation (one
Pomeron splitting into two) and recombination (two Pomerons
merging into one). It is constructed by exploiting a duality
principle relating the evolutions in the target and the
projectile, more precisely, splitting and merging processes, or
fluctuations in the
dilute regime and saturation effects in the dense regime. The
simplest Pomeron loop calculated with the effective theory is free
of both ultraviolet or infrared singularities.

\end{abstract}
\end{center}}

\end{frontmatter}
\newpage
There has been recently significant progress in our understanding of
high energy hadronic scattering, and in particular of the
processes occurring at large parton densities and which are
believed to be responsible for the unitarization of the scattering
amplitudes and the saturation of the parton distributions. Non linear
evolution equations have been derived which describe the
approach towards saturation and the unitarity limit, and
which have the structure of stochastic evolution equations.
However, it has been very recently recognized \cite{IT04} that the
equations which were considered as the most complete, namely
the Balitsky--JIMWLK
(Jalilian-Marian--Iancu--McLerran--Weigert--Leonidov--Kovner) equations
\cite{B,JKLW,RGE,W}, are in fact incomplete. This
is manifest in the statistical language by the presence of
fluctuations at high momenta \cite{MS04,IMM04} which are not well
accounted for by the JIMWLK evolution of the target wavefunction \cite{IT04}.
In the language of Pomerons, the JIMWLK equation contains Pomeron
merging but not also Pomeron splitting\footnote{Note that
the opposite terminology for what one means by `splitting' and `merging'
would be more natural in relation
with Balitsky equations, which refer to the evolution of the projectile.
To avoid confusion on this point, in this paper we shall systematically
use the terminology appropriate to target evolution.}.

Following this observation, two of us (E.I. and D.T.) have constructed
a hierarchy of non--linear evolution equations for the dipole
scattering amplitudes which include both gluon mergings and gluon
splittings, and thus generate Pomeron loops through iterations
\cite{IT04,IT05}. These equations have been argued to hold in the limit
where the number of colors $N_c$ is large ($N_c\gg 1$), and indeed it
has been checked explicitly in Ref. \cite{IT05} that the vertices
appearing in these equations are the same as the corresponding `triple
Pomeron vertices' computed in perturbative QCD at large $N_c$
\cite{BW93,BV99,BLV05}. A complementary approach has been developed by
Mueller, Shoshi, and Wong  \cite{MSW05} who proposed a generalization
of the JIMWLK equation which includes the effects of pomeron splitting
in the dilute regime and for large $N_c$. These two approaches follow
the same general strategy --- namely, they combine the non--linear
JIMWLK equation at high density with the color dipole picture
\cite{AM94,AM95} in the dilute regime --- and lead indeed to the same
evolution equations for the scattering amplitudes, as demonstrated in
Ref. \cite{IT05}. (See also Refs. \cite{LL05,KL05} for related recent
developments.)

It is our purpose in this letter to show that the equations
obtained  in \cite{IT04,IT05} and \cite{MSW05} can be reformulated
in term of an effective theory for Pomerons. By 'Pomeron' we mean
here the color singlet exchange which describes the interaction
between an elementary color dipole and the field of a target in a
single scattering approximation, and which reduces to two gluon
exchanges in lowest order perturbation theory. The construction of
the effective theory involves a projection onto restricted degrees
of freedom, precisely the Pomerons, and is expressed in terms of a
simple Hamiltonian which describes the BFKL evolution of the
Pomerons together with their splitting and merging. By requiring
that the evolution should lead to identical results whether it is
viewed as the evolution of the target or that of the projectile,
one arrives at a duality principle which is used to construct the
effective Hamiltonian from the Hamiltonian derived in \cite{MSW05}
in the dilute regime. The limitations of the effective theory, and
the subtle mathematical problems that arises when one attempts to
analyze its microscopic content will be briefly discussed at the
end of this letter.

Most treatments of high energy scattering rely on an asymmetric
approach: typically, the `projectile' is viewed as a
collection of test particles which  probe the color field of the
`target'. At high energy, the eikonal approximation is a good
approximation, and the scattering of an elementary color charge is
described by a {\em `Wilson line'} of the form
 \begin{equation}
 \label{Vdef}
  V^\dagger_{\bm{x}}[\alpha] \,\equiv\,{\rm P}\,{\rm
  exp}\left(i g\int dx^- \alpha^a(x^-,{\bm{x}}) t^a\right),
 \end{equation}
where ${\bm{x}}$ denotes the transverse coordinate of the
particle, which is not affected by its interactions with the
field of the target $\alpha^a(x^-,{\bm{x}})$, $t^a$ are the generators
of the SU(3) algebra in the representation appropriate for the
test particle,  and the symbol P indicates that, in the expansion
of the exponential, the color matrices $\alpha^a(x^-,{\bm{x}})
t^a$ must be ordered from right to left in increasing order  in
$x^-$ (we are using light--cone vector notations, $x^\pm \equiv
(t\pm z)/\sqrt{2}$.). For a more complex projectile, viewed as a
collection of elementary color charges, the $S$--matrix is given
by a product of Wilson lines like Eq.~(\ref{Vdef}), one for each
elementary color charge.

In a frame in which most of the total rapidity $Y$ is carried by
the target, the target wavefunction  can be described as a {\em
color glass condensate} \cite{MV,RGE}, and the corresponding
$S$--matrix is obtained as:
 \be \lan S\ran_Y =  \int
 {\rm D}[\alpha]\,W_Y[\alpha]\,S[\alpha]\,,
 \label{Sgen} \ee
where $\alpha\equiv \alpha^a(x^-,{\bm{x}})$  is a classical field
randomly distributed with {\em weight function} $W[\alpha]$ (a
functional probability distribution), and $S[\alpha]$ is the
projectile $S$--matrix for a given configuration of this random
field. With increasing $Y$, the weight function evolves according
to a functional renormalization group equation, of the generic
form:
 \be\label{RGEgen}
 \frac{\partial}{\partial Y} \,W_Y[\alpha]= -
     H\Big[\alpha, \frac{\delta}{i\delta \alpha}\Big]
     W_Y[\alpha]\,,
  \ee
where $H$ is a functional differential operator  commonly referred
to as the  `Hamiltonian'.   Alternatively, one can view the same
evolution as a change in the scattering operator, for a fixed
weight function $W[\alpha]$.
 To see that, take a derivative w.r.t. $Y$
in Eq.~(\ref{Sgen}), use Eq.~(\ref{RGEgen}), and perform an integration by part in the functional integral:
 \be\label{DSgen}
 \frac{\partial}{\partial Y} \,\lan S\ran_Y =  - \int
 {\rm D}[\alpha]\,W[\alpha]\, H^\dagger\Big[\alpha,
  \frac{\delta}{i\delta \alpha}\Big]\,S[\alpha]\,.\ee
This can be interpreted as describing the evolution of the
scattering operator $S_Y[\alpha]$, with `Hamiltonian' $H^\dagger$:
  \be\label{HSgen}
 \frac{\partial}{\partial Y} \, S_Y[\alpha]= -
     H^\dagger\Big[\alpha, \frac{\delta}{i\delta \alpha}\Big]
     S_Y[\alpha]\,.
  \ee
Both points of view, somewhat reminiscent of, respectively, the
Schr\"odinger and the Heisenberg pictures of quantum mechanics, will
be used in the following discussion (although we shall refrain from
introducing explicitly rapidity dependent operators). In the
Schr\"odinger picture, one puts emphasis on the evolution of the
state vector, whose role is played here by the weight functional
$W_Y[\alpha]$. In the Heisenberg picture, the state vector is a
constant reference vector involved in the calculation of all
expectation values, here $W[\alpha]$, and one puts all the evolution
in the operators, here the scattering operators $S_Y[\alpha]$. The
Schr\"odinger picture corresponds to evolution equations which aim
at providing a detailed microscopic description of the color field
in the target, together with its complicated correlations. This is
what the JIMWLK equation does. The Heisenberg picture rather
describes how the test particles get dressed by color field
fluctuations as they are boosted to higher rapidities. In this
approach, the complicated color correlations in the target
wavefunction are not immediately visible, and indeed the resulting
equation of motion are established somewhat more easily.  This
second approach is essentially the one used by Balitsky to obtain
his hierarchy of equations.

The test particles that we shall consider are in fact elementary
color dipoles, whose scattering amplitude reads:
 \be T(\x,\y)= 1 -
 \frac{1}{N_c}\tr(V^\dag_{\x} V_{\y}),\label{Tdipole} \ee
for a dipole with the quark leg at $\x$ and the antiquark leg at $\y$.
Here the
Wilson lines are taken in the fundamental representation. We shall
be interested in situations where the dipoles scatter off the
color glass in the two-gluon exchange approximation (weak field
limit) and we shall work in a large-$N_c$ limit. In the weak field
limit, the amplitude for a single dipole to scatter is obtained
after expanding each of the Wilson lines to second order in
$\alpha$:
   \be \label{Vexp}\hspace*{-5mm}
V^\dagger_{\bm{x}}[\alpha] &= &1\,+\, ig\int dx^-
\alpha^a(x^-,{\bm{x}}) t^a\nn &-&\frac{g^2}{2} \int dx^-\!\int dy^-
\alpha^a(x^-,{\bm{x}})\alpha^b(y^-,{\bm{x}})\big[\theta(x^- -
y^-)t^a
 t^b + \theta(y^- - x^-)t^b t^a\big]\nonumber\\
  &+&\quad\cdots \,\,.\ee
Note that, to this order, the $x^-$--ordering of the color
matrices starts to play a role in Eq.~(\ref{Vexp}). Still, this
ordering is irrelevant for the computation of the dipole amplitude
to lowest order, because of the symmetry of the color trace: ${\rm
tr} (t^a t^b)=\frac{1}{2}\,\delta^{ab}={\rm tr} (t^b t^a)$.
Namely, one finds:
\begin{align}\label{Tdipole0}
    T(\x,\y) \simeq T_0(\x,\y)\equiv
    \frac{g^2}{4 N_c}
    \left[\alpha^a({\x})-\alpha^a({\y})\right]^2,
\end{align}
which involves only the integrated field $\alpha^a(\x)\equiv \int
dx^- \alpha^a(x^-,{\bm{x}})$. Similarly the amplitude for $\kappa$
dipoles to scatter is given, within the same approximation, by
$T_0^{(\kappa)}(\x_1,\y_1,...,\x_{\kappa},\y_{\kappa})=T_0(\x_1,\y_1)
... T_0(\x_{\kappa},\y_{\kappa})$. { In what follows, we shall
refer to the  amplitude (\ref{Tdipole0}) describing the
single--scattering of an elementary dipole off a given color field
as to a ``Pomeron exchange''. Similarly, $T_0^{(\kappa)}$ describes
the exchange of $\kappa$ Pomerons.}

At this point we find it useful to digress on the linear evolution
equation known as BFKL equation. This will allow a few observations
which illuminate some of the mathematical subtleties involved in
taking the large $N_c$ limit when constructing our effective theory.
Consider first the JIMWLK Hamiltonian. As shown in
Ref.~\cite{ODDERON}, when it is restricted to act on gauge-invariant
observables, it can be given the simple form:
\begin{align}\label{HJIMWLK}
 H_{\rm JIMWLK} =
 -\frac{1}{16\pi^3} \int\limits_{\bm{u},\bm{v},\bm{z}}&
 {\mathcal M}({\bm{u}, \bm{v}, \bm{z}})
 \left(1+\V^\dag_{\uu}
 \V_{\bm{v}} -\V^\dag_{\uu} \V_{\z} - \V^\dag_{\z} \V_{\bm{v}} \right)^{ab}
 \nonumber \\
 &\times
 \frac{\delta}{i\delta\alpha^a_Y(\uu)}
 \frac{\delta}{i\delta\alpha_Y^b(\bm{v})},
 \end{align}
where $\mathcal{M}$ is the dipole kernel
\begin{align}\label{kernel}
    \cal{M}(\bm{x},\bm{y},\bm{z}) =
    \dk{\bm{x}}{\bm{y}}{\bm{z}}.
\end{align}
Here the Wilson lines are in the adjoint representation. The
derivatives can be freely moved across the bilinear form in Wilson
lines, because they commute with the latter in the presence of the
dipole kernel. That is, $H_{\rm JIMWLK}$ is Hermitian. The BFKL
limit of $H_{\rm JIMWLK}$ is obtained by expanding the Wilson lines
to lowest non-trivial order in $\alpha$. One gets:
 \begin{align}\label{weak}
 H_{\rm BFKL} = -\,\frac{g^2}{16\pi^3}
 \int\limits_{\bm{u},\bm{v},\bm{z}}
 &{\mathcal  M}({\bm{u}, \bm{v}, \bm{z}})
 [\alpha^a(\bm{u})-\alpha^a(\bm{z})]
 [\alpha^b(\bm{v})-\alpha^b(\bm{z})]
\nonumber \\
 &\times f^{acf} f^{bfd}
 \frac{\delta}{\delta\alpha^c_Y(\uu)}\frac{\delta}
 {\delta\alpha_Y^d(\bm{v})}.\,\,
\end{align}
It is not difficult to verify that $H_{\rm BFKL}$ is again
Hermitian.

Let us now turn to the `large--$N_c$ limit'. This is obtained by
{\sf (i)} restricting the action of $H_{\rm BFKL}$ to the dipole
operators $T_0^{(\kappa)}$ mentioned above and {\sf (ii)} preserving
only the dominant terms at large $N_c$ in the action of the Hamiltonian
on these operators. When acting on the
color fields inside a single factor $T_0$ (i.e., on the same
dipole), the two functional derivatives in $H_{\rm BFKL}$ yield a
factor $\delta^{cd}$, and then $f^{acf} f^{bfc} = -N_c\delta^{ab}$
produces the expected $N_c$ enhancement. On the other hand, the
action  on the color fields within two different factors $T_0$
(i.e., upon two different dipoles) produces no such enhancement.
Thus, at large $N_c$, $H_{\rm BFKL}$ can be equivalently replaced
by an effective Hamiltonian in which the two functional
derivatives are traced over color. This Hamiltonian, which we
denote $H_0^\dagger$ for reason which will become clear shortly,
is
\begin{align}\label{HBFKL}
    H_0^\dagger =
    \frac{1}{2 N_c^2}\, \atpi
    \int\limits_{\bm{u},\bm{v},\bm{z}}
    \cal{M}(\bm{u},\bm{v},\bm{z})
    \left[ \alpha^a(\bm{u}) -\alpha^a(\bm{z}) \right]
    [\alpha^a(\bm{v}) -\alpha^a(\bm{z})]
    \frac{\delta}{\delta \alpha^b(\bm{u})}
    \frac{\delta}{\delta \alpha^b(\bm{v})},
\end{align}
where  $\abar=\alpha_s N_c/\pi$. Let us emphasize that, as obvious
from the construction we have given, the two derivatives in
$H_0^\dagger$ are to act on {\em the same dipole}. Note also that,
as opposed to the original $H_{\rm BFKL}$, $H_0^\dagger$ is not
Hermitian: in fact, it is readily seen that its adjoint is
ill-defined. This reflects the fact that the construction of
$H_0^\dagger$ involves a projection on a specific set of degrees
of freedom, and once this is done, one looses the possibility to
integrate by part as in Eq.~(\ref{DSgen}) in order to let $H_0$
act on the weight functional $W[\alpha]$. These special
mathematical properties, restriction of the space on which the
Hamiltonian is acting  and loss of hermiticity, are general, and
peculiar, mathematical features of the effective theory that we
shall present. It is tempting to speculate  that in doing  the
large $N_c$ limit we are renouncing to follow the evolution of
some color correlations (precisely those which are suppressed at
large $N_c$). The corresponding loss of information may be
responsible for the simpler Markovian stochastic theory that we
shall arrive at.

We now return to the main stream of our discussion and establish a
useful
 property. In Ref. \cite{IM031}, a symmetric description
was obtained for the scattering between two
 color glasses in the regime where {\em both} systems are in the weak
field regime). The final formula reads
  \be\label{SYCGC} \lan S\ran_Y
\,=\,\int {\rm D}[\alpha_R]\, \,W_{Y-y}[\alpha_R]\int {\rm
D}[\alpha_L]\, \,W_y[\alpha_L] \,\,{\rm e}^{\,i\int d^2{\bm
 z}\,\rho^a_L({\bm z}) \alpha^a_R({\bm{z}})}\,.\ee
In this expression, $\rho_L^a(\x) =
-\nabla^2_{\bm{x}}\alpha_L^a(\x)$ is the classical color charge
density of the left--mover, and $W_{Y-y}[\alpha_R]$ and
$W_y[\alpha_L]$ are the weight functions for the right--moving and,
respectively, left--moving color glass (note that the rapidity of
the left mover is measured positively to the left, so that as we
vary $y$, the total rapidity interval between projectile and target
remains equal to $Y$). The precise conditions for the validity of
Eq.~(\ref{SYCGC}) are detailed in Ref. \cite{IM031}. Let us
emphasize here a non-trivial aspect of this formula. Although it is
essentially a weak field formula which assumes that the elementary
dipoles interact only once, it contains the possibility that any
number of dipoles of the projectile interact with an equivalent
number of dipoles in the target. Thus Eq.~(\ref{SYCGC}) does account
for multiple scattering, albeit in a restrictive way (each dipole
interacting only once). These multiple scattering generate unitarity
corrections if  $Y$ is large enough. At the same time, we require
both color glasses to be unsaturated. This  imposes some limited
range of variation for $y$ within which Eq.~(\ref{SYCGC}) is
correct.

Now, Lorentz invariance implies that $\langle S_Y\rangle$ may
depend on the total rapidity interval $Y$, but, within the range
of validity of Eq.~(\ref{SYCGC}), cannot depend upon the rapidity
$y$ used to separate the system into a `projectile' and a
`target', or equivalently on the frame which we choose to describe
the collision. {
This implies (see also Ref. \cite{KL3} for a similar argument):
\be\label{DSYCGC} 0\,=\, \frac{\partial \langle S\rangle_Y}{\partial y}
 &=&\int {\rm D}[\alpha_R]\, \int {\rm D}[\alpha_L]\,\,{\rm e}^{\,i \int d^2{\bm
 z}\,\rho^a_L({\bm z}) \alpha^a_R({\bm{z}})}\nn &{}&\ \ \left\{\Big(
 \frac{\partial}{\partial y} W_{Y-y}[\alpha_R]\Big)W_y[\alpha_L]
 +W_{Y-y}[\alpha_R] \Big( \frac{\partial}{\partial y}
 \,W_y[\alpha_L]\Big)\right\} \,\,.\ee
The evolution of both weight functions are given by:
 \be  \label{DWW}
     \frac{\partial}{\partial y} \,W_{Y-y}[\alpha_R] &= &-
     \frac{\partial}{\partial Y} \,W_{Y-y}[\alpha_R] =
     H\Big[\alpha_R, \frac{\delta}{i\delta \alpha_R}\Big]
     W_{Y-y}[\alpha_R],\nn
    \frac{\partial}{\partial y} \,W_y[\alpha_L]&= &-
     H\Big[\alpha_L, \frac{\delta}{i\delta \alpha_L}\Big]
     W_y[\alpha_L].
  \ee
We shall keep the evolution of the left--mover as shown in the above
equation, but  perform an integration by parts in the
functional integral over $\alpha_R$ in Eq.~(\ref{DSYCGC}). Next, we note that
 \be\hspace*{-3mm}\label{predual}
 H^\dagger\Big[\alpha_R, \frac{\delta}{i\delta \alpha_R}\Big]
 \,{\rm e}^{\,i\int d^2{\bm
 z}\,\rho^a_L({\bm z}) \alpha^a_R({\bm{z}})}\,=\,
 H\Big[\frac{\delta}{i\delta \rho_L},
 \rho_L\Big]
 \,{\rm e}^{\,i\int d^2{\bm
 z}\,\rho^a_L({\bm z}) \alpha^a_R({\bm{z}})}.
  \ee
Using this identity in  Eq.~(\ref{DSYCGC}) and performing
 a further integration by parts, now w.r.t. $\alpha_L$ (recall that
$\rho_L^a(\x) = -\nabla^2_{\bm{x}}\alpha_L^a(\x)$), one is left with
a differential operator acting on $W_y[\alpha_L]\equiv W_y[\rho_L]$
(with a slight abuse in the notation):\be
 H^\dagger\Big[ \frac{\delta}{i\delta \rho_L},
 \rho_L\Big]\, W_y[\rho_L].\ee
For Eq.~(\ref{DSYCGC}) to be satisfied, the contribution above
should cancel against the term in Eq.~(\ref{DWW}) describing the
evolution of the left--mover. This condition leads to the
`self-duality' condition:
 \be\label{DUAL}
 H\Big[\alpha_L, \frac{\delta}{i\delta \alpha_L}\Big]
     W_y[\alpha_L]\,=\,H^\dagger\Big[\frac{\delta}{i\delta \rho_L},
 \rho_L\Big]\, W_y[\rho_L]\,.
  \ee
  The same relation holds obviously for the
`right' variables $\alpha_R,\rho_R$.}

\begin{figure}[htb]
\begin{center}
\includegraphics[scale=1.05]{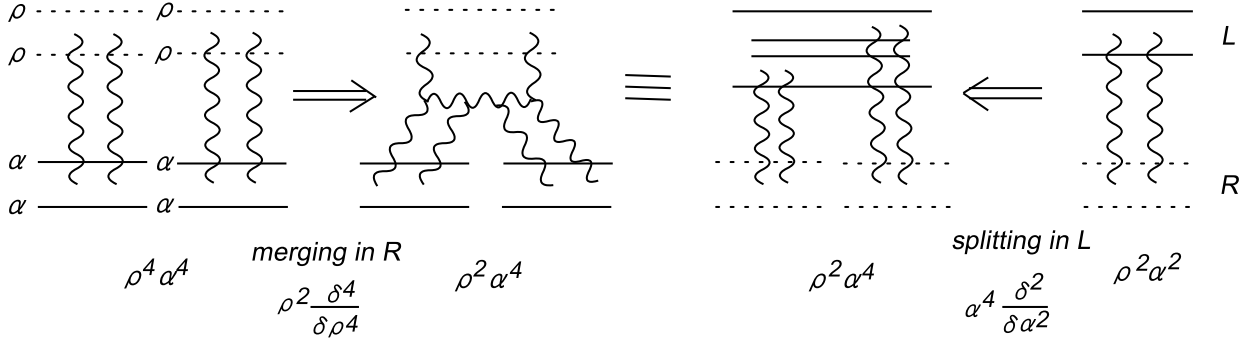}
    \caption{An illustration of the relation (\ref{predual}). $\rho$ denotes
the color charge of the left--mover (L) and $\alpha$ is the color
field of the right--mover (R). The same diagram (any of the two diagrams
in the middle) can be viewed either as a merging in R, or as a splitting in L.
The first interpretation is natural when the diagram is produced
by acting on the eikonal line with the Hamiltonian $H^\dagger_{1\to 2}\sim
\rho^2\delta^4/\delta \rho^4$ for L (cf. Eq.~(\ref{MSW})). The second
interpretation rather corresponds to the action of
$H^\dagger_{2\to 1}\sim \alpha^4\delta^2/\delta \alpha^2$ for R
(cf. Eq.~(\ref{H42}).}
\label{dualfig}
\end{center}
\end{figure}

Going back to Eq.~(\ref{predual}), one sees that what is involved in
the duality operation\footnote{To our knowledge, the duality between the
roles of the operators $\rho^2\delta^n/\delta \rho^n$ and
$\alpha^n\delta^2/\delta \alpha^2$ has been first recognized by L.
McLerran.} is a matching of splitting processes in the
left movers, encoded by terms in the Hamiltonian of the form
$\rho^2\delta^n/\delta \rho^n$, into merging process in the right
movers, corresponding to terms of the form $\alpha^n\delta^2/\delta
\alpha^2$. An example of such a matching is illustrated in Fig. \ref{dualfig}.
 Splitting terms dominate in the dilute regime where
they control the fluctuations, while merging terms become essential in
the saturation regime where parton densities are large. This
{\it fluctuation--saturation duality} is turned into a constraint on the
evolution Hamiltonian of either the projectile or the target in
Eq.~(\ref{DUAL}).

{ The self-duality constraint, which we expect to hold within
the limited range of energies in which the factorization
(\ref{SYCGC}) is valid\footnote{The self-duality condition (\ref{DWW}) has
recently been claimed to hold in a much broader context \cite{KL3}.}
 \cite{IM031}, will be used now to construct
a simple Hamiltonian  describing the Pomeron dynamics in the
dilute regime, starting from the known, dominant, contribution
containing only splitting processes that has been constructed by
Mueller, Shoshi and Wong \cite{MSW05}. Quite remarkably, and
somewhat unexpectedly, this Hamiltonian leads to equations of
motion which  reproduces the exact ones at large $N_c$
\cite{IT04,IT05}, that is the effective theory appears to be valid
beyond the dilute limit where it is established. The Hamiltonian
constructed in  \cite{MSW05} reads}
\begin{align}\label{MSW}
    H_{1 \to 2} =
    -\frac{g^2}{16 N_c^3}\, \atpi
    \int & \cal{M}(\bm{u},\bm{v},\bm{z})
    \cal{G}(\bm{u}_1|\bm{u},\bm{z})
     \cal{G}(\bm{v}_1|\bm{u},\bm{z})
    \cal{G}(\bm{u}_2|\bm{z},\bm{v})
    \cal{G}(\bm{v}_2|\bm{z},\bm{v})
    \nonumber\\
    & \times
    \frac{\delta}{\delta \alpha^a(\bm{u}_1)}
    \frac{\delta}{\delta \alpha^a(\bm{v}_1)}
    \frac{\delta}{\delta \alpha^b(\bm{u}_2)}
    \frac{\delta}{\delta \alpha^b(\bm{v}_2)}
    \nabla_{\bm{u}}^2 \nabla_{\bm{v}}^2
    \alpha^c(\bm{u}) \alpha^c(\bm{v}).
\end{align}
In Eq.~(\ref{MSW}),  the
integration goes over all the transverse coordinates $\bm{u}$,
$\bm{v}$, $\bm{z}$, $\bm{u}_1$, $\bm{v}_1$, $\bm{u}_2$,
$\bm{v}_2$. The function $\cal{G}(\bm{u}_1|\bm{u},\bm{z})$ is, up
to a factor $g\,t^a$, the classical field created at $\bm{u}_1$ by
the elementary dipole $(\bm{u},\bm{z})$, and reads
\begin{align}\label{calG}
    \cal{G}(\bm{u}_1|\bm{u},\bm{z}) =
    \frac{1}{4\pi}
    \ln \frac{(\bm{u}_1-\bm{z})^2}{(\bm{u}_1-\bm{u})^2}.
\end{align}
It is  easy to understand (and was explicitly shown in
\cite{IT05}) that this Hamiltonian generates {\it Pomeron
splittings}. More precisely, the result of the operation of
$H^\dagger_{1 \to 2}$ on the two--Pomeron exchange amplitude
$T^{(2)}_0$ is proportional to ${T_0}$, and thus generates the
following, {\it fluctuation}, term in the evolution equation for
$T_0^{(2)}(\bm{x}_1,\bm{y}_1;\bm{x}_2,\bm{y}_2)$:
\begin{align}\label{T2evol}
    H^{\dag}_{1 \to 2} T_0^{(2)}=
    \left(\frac{\alpha_s}{2\pi}\right)^2
    \frac{\abar}{2 \pi}\!
    \int\limits_{\bm{u},\bm{v},\bm{w}}&
    \mathcal{M}(\bm{u},\bm{v},\bm{w})\,
    \cal{A}_0(\bm{x}_1,\bm{y}_1|\bm{u},\bm{w})\,
    \cal{A}_0(\bm{x}_2,\bm{y}_2|\bm{w},\bm{v})\,
    \nabla_{\bm{u}}^2 \nabla_{\bm{v}}^2\, T_{0}(\bm{u},\bm{v}),
\end{align}
where $\alpha_s^2\cal{A}_0$ is the amplitude for dipole--dipole scattering
in the two--gluon exchange approximation and for large $N_c$ :
\begin{align}\label{T0}
    \cal{A}_0(\bm{x},\bm{y}|\bm{u},\bm{v}) =
    \frac{1}{8}
    \left[\ln \frac{(\bm{x}-\bm{v})^2 (\bm{y}-\bm{u})^2}
    {(\bm{x}-\bm{u})^2 (\bm{y}-\bm{v})^2}
    \right]^2.
\end{align}
Clearly, this process corresponds to the splitting of one Pomeron
into two. In general the Hamiltonian $H_{1 \to 2}$ can describe
the transition $n \to n+1$, in which case  $n-1$ of the Pomerons
are simply ``spectators''. Note that  $H_{1 \to 2}$ is
non-Hermitian, which we interpret as reflecting again the
large--$N_c$ approximation implicitly involved in its derivation.

To apply the duality transformation, it is convenient to reexpress
$H_{1 \to 2}$  in terms of the sources $\rho^a(\x)$ of the color
field $\alpha^a(\bm{x})$, by using $\rho^a(\x) =
-\nabla^2_{\bm{x}}\alpha^a(\x)$. We then obtain \be\label{H24}
    H_{1 \to 2}&=&
    -\frac{g^2}{16 N_c^3}\, \atpi
    \int\limits_{\bm{u},\bm{v},\bm{z}}
    \cal{M}(\bm{u},\bm{v},\bm{z})\nn
   &{}&\qquad\times\
  \left[ \frac{\delta}{\delta \rho^a(\bm{u})} -
           \frac{\delta}{\delta \rho^a(\bm{z})} \right]^2
    \left[ \frac{\delta}{\delta \rho^b(\bm{z})} -
           \frac{\delta}{\delta \rho^b(\bm{v})} \right]^2
    \rho^c(\bm{u}) \rho^c(\bm{v}).
\ee
At this point we force the Hamiltonian to be self-dual.
This is done by adding to
$H_{1 \to 2}[{\delta}/{i\delta \rho},\rho] $ its dual
$H^\dagger_{1 \to 2}[\alpha, {\delta}/{i\delta \alpha}]
 \equiv H_{2 \to 1}$ (this new notation will be justified shortly).
 The Hermitian conjugate of $H_{2 \to 1}$ reads
\begin{align}\label{H42}
    H_{2 \to 1}^\dagger=
    \frac{g^2}{16 N_c^3}\, \atpi
    \int\limits_{\bm{u},\bm{v},\bm{z}}
    \cal{M}(\bm{u},\bm{v},\bm{z})
    \left[ \alpha^a(\bm{u}) -\alpha^a(\bm{z}) \right]^2
    [\alpha^b(\bm{z}) -\alpha^b(\bm{v})]^2
    \frac{\delta}{\delta \alpha^c(\bm{u})}
    \frac{\delta}{\delta \alpha^c(\bm{v})},
\end{align}
and the action of $H_{2 \to 1}^\dagger$ on the dipole scattering
amplitude is
\begin{align}\label{21T0}
    H^\dagger_{2 \to 1}\,T_0(\bm{x},\bm{y})
    &=\atpi \int\limits_{\bm{z}}
    \cal{M}(\bm{x},\bm{y},\bm{z})\,
    \frac{g^4}{16 N_c^2}\,
    [\alpha^{a}(\bm{x}) - \alpha^{a}(\bm{z})]^2
    [\alpha^{b}(\bm{z}) - \alpha^{b}(\bm{y})]^2
    \nonumber \\
    &=\atpi \int\limits_{\bm{z}}
    \cal{M}(\bm{x},\bm{y},\bm{z})\,
    T^{(2)}_0(\bm{x},\bm{z};\bm{z},\bm{y}).
\end{align}
Thus $H_{2 \to 1}$ generates the non--linear term in the first
Balitsky equation. Similarly, it is obvious to show that the
operation on $T^{(\kappa)}_0
(\bm{x}_1,\bm{y}_1;...;\bm{x}_{\kappa},\bm{y}_{\kappa})$, will
generate correctly the non-linear terms of the $\kappa$--the
Balitsky equation in the large--$N_c$ limit (this is trivial; only
one amplitude is ``active'', and we need to take into account all
the possible permutations). Therefore the Hamiltonian in
Eq.~(\ref{H42}) generates in  an effective way {\it Pomeron
mergings} (hence the notation $H_{2 \to 1}$); one has a transition
of the form $n+1 \to n$ where, again, $n-1$ of the Pomerons are
spectators.

Thus the total Hamiltonian of our  Pomeron effective theory reads
\begin{align}\label{Htotal}
    H^\dagger=  H_0^\dagger+ H_{1 \to 2}^\dagger + H_{2 \to 1}^\dagger.
\end{align}
The  Hamiltonian $H_0^\dagger$, describing the BFKL evolution, plays
here the role of the free Pomeron Hamiltonian. The other two pieces
$H^\dagger_{2\to 1}$ and $H^\dagger_{1 \to 2}$ correspond
respectively  to Pomeron merging and splitting, and will naturally
generate {\it Pomeron loops} in the course of the evolution. The
minimal Pomeron loop, which is simply the one--loop correction to
the scattering amplitude $\mean{T(\bm{x},\bm{y})}$, can be isolated
by the successive operation of these two parts of the Hamiltonian,
namely ${\mathbb P}{\mathbb L} = H^\dagger_{1\to 2} H^\dagger_{2 \to 1} T_0$. The
explicit results reads
 \begin{align}\label{PL}
    {\mathbb P}{\mathbb L} = -
    \left(\frac{\abar}{2\pi}\right)^2 \left(\frac{\alpha_s}{2\pi}\right)^2
    \int\limits_{\bm{u},\bm{v},\bm{z},\bm{w}}
&   \cal{M}(\bm{x},\bm{y},\bm{z})\cal{M}(\bm{u},\bm{v},\bm{w})\\
&   \hspace*{-5mm}\times \cal{A}_0(\bm{x},\bm{z}|\bm{u},\bm{w})
    \cal{A}_0(\bm{z},\bm{y}|\bm{w},\bm{v})
\nabla_{\bm{u}}^2 \nabla_{\bm{v}}^2\,
\mean{T_0(\bm{u},\bm{v})}.\nonumber
\end{align}
Note that this result is free of any (ultraviolet or infrared)
divergences. For instance, the poles in the dipole kernel at
$\bm{z}=\bm{x}$ and $\bm{y}$ are harmless because of
$\cal{A}_0(\bm{x},\bm{z}=\bm{x}|\bm{u},\bm{w})=0$.

A simple physical picture of this result is obtained by assuming
that this Pomeron loop has been generated after the first two
steps in the evolution starting with a target which is itself an
elementary dipole $(\bm{x_0},\bm{y_0})$. Then, Eq.~(\ref{PL})
simplifies to:
 \begin{align}\label{PL0}
    {\mathbb P}{\mathbb L}^0 = -2
    \left(\frac{\abar}{2\pi}\right)^2 \alpha_s^4
    \int\limits_{\bm{z},\bm{w}}
\cal{M}(\bm{x},\bm{y},\bm{z})\cal{M}(\bm{x_0},\bm{y_0},\bm{w})
\cal{A}_0(\bm{x},\bm{z}|\bm{x_0},\bm{w})
    \cal{A}_0(\bm{z},\bm{y}|\bm{w},\bm{y_0}).
\end{align}
This result has a clear physical interpretation: Both original
dipoles --- in the projectile and the target --- split into new
dipoles, processes which are represented by the two dipole kernels
times $\bar\alpha_s^2$. Then, the child dipoles from the two
systems scatter with each other, by exchanging two pairs of
gluons; this yields the two factors ${\cal A}_0$ times
$\alpha_s^4$. Finally, note that this contribution is negative, as
expected, leading to a decrease in the amplitude in the course of
the evolution.

As we have already emphasized, the  Hamiltonian (\ref{Htotal})
reproduces the complete equations of motion established in
\cite{IT04,IT05} and \cite{MSW05}. While this intriguing property
deserves further investigation, some insight can be gained by
analyzing how the merging processes in the effective Hamiltonian
compare to those deduced from correct microscopic dynamics as
described by JIMWLK. The action of $H_{\rm JIMWLK}$,
Eq.~(\ref{HJIMWLK}), on the full dipole scattering amplitude
$T(\x,\y)$, Eq.~(\ref{Tdipole}), is
\begin{align}\label{func_der_full}
    \frac{\delta}{\delta \alpha^a(\bm{u})}
    \frac{\delta}{\delta \alpha^b(\bm{v})}
  T(\x,\y)=
    \frac{g^2}{N_c}
    (\delta_{\bm{y}\bm{v}} - \delta_{\bm{x}\bm{v}})
    \left[ \delta_{\bm{u}\bm{x}}\,
    \tr (t^b t^a V^\dag_{\bm{x}} V_{\bm{y}})
    - \delta_{\bm{u}\bm{y}}\,
    \tr(t^a t^b V^\dag_{\bm{x}} V_{\bm{y}}) \right]
\end{align}
Simple algebra then easily yields the first Balitsky equation :
\begin{align}\label{T1evolfull}\hspace*{-5mm}
    H_{\rm JIMWLK} T(\bm{x},\bm{y}) =
    \frac{\abar}{2\pi} \int\limits_{\bm{z}}
    \cal{M}(\bm{x},\bm{y},\bm{z})
    [-T(\bm{x},\bm{y})+T(\bm{x},\bm{z})+
    T(\bm{z},\bm{y}) -\,T(\bm{x},\bm{z})T(\bm{z},\bm{y})].
\end{align}
Then, after expanding the dipole operator $T$ in the weak-field
limit, and keeping terms up to the quartic order with respect to
gauge field $\alpha$, one finds an evolution equation which contains
not only the BFKL dynamics, but also the lowest order   mergings
(four gluons merging into two).

Consider now the action of  the JIMWLK Hamiltonian on
$T_0(\x,\y)$. Since:
\begin{align}\label{func_der_dipole}
    \frac{\delta}{\delta \alpha^a(\bm{u})}
    \frac{\delta}{\delta \alpha^b(\bm{v})}\,
    T_0(\x,\y) = \frac{g^2}{2 N_c}\,
    \delta^{ab} (\delta_{\x\bm{u}}-\delta_{\y\bm{u}})
    (\delta_{\x\bm{v}}-\delta_{\y\bm{v}}),
\end{align}
we have:
\begin{align}\label{dipole_ev}
    H_{\rm JIMWLK} T_0(\x,\y)
    =\frac{g^2}{2 N_c^2}\, \atpi
    \int\limits_{\z} {\cal M}(\x,\y,\z)
    \Tr
    \left(1+ \V^\dag_{\x} \V_{\y}-
    \V^\dag_{\x} \V_{\z} - \V^\dag_{\z} \V_{\y}
    \right).
\end{align}
 When expanding the Wilson lines in the r.h.s.
in powers of $\alpha$, one obtains quadratic terms describing the BFKL
evolution of ${T_0(\x,\y)}$ plus higher order terms which describe
$n\to 2$ gluon mergings. But at this level, it is easy to see that the
$4\to 2$ terms generated by this expansion are not the same as
those in the r.h.s. of Eq.~(\ref{21T0}). For instance, while
the merging term in Eq.~(\ref{21T0}) includes a piece
containing three different transverse positions (i.e.,
$\alpha_{\x}^a \alpha_{\z}^a \alpha_{\z}^b \alpha_{\y}^b$), the
corresponding JIMWLK result in Eq.~(\ref{dipole_ev})
cannot generate such terms.

We thus see that thet actual, microscopic,
dynamics of gluon merging in QCD is considerably more complicated than
in our simple effective theory, yet the latter provides, as we have
seen, the correct evolution equations for the scattering amplitudes.
This shows that the additional merging terms generated by the JIMWLK
Hamiltonian must compensate in the evolution equations against non--linear
(quartic in $\alpha$, or higher) contributions to the scattering amplitudes,
as obtained by expanding the Wilson lines in equations like (\ref{Tdipole}).
We can refer to the latter as describing the {\it dressing of the pomeron
with multiple scattering}.

This brings us to comment on  the nature of the dynamics described
by the effective theory. This theory generates evolution equations
for the Pomeron operators $T_0^{(\kappa)}$ which are formally
identical to the equations satisfied by the complete dipole
scattering operators $T^{(\kappa)}$ in QCD at large $N_c$. This
means in particular that the solutions to the equations for $\lan
T_0^{(\kappa)}\ran_Y$ will appear to saturate the unitarity (or
`black disk') limit $T_0=1$ in the high energy limit, in spite of
the fact that the respective operators describe single scatterings
only ! This indicates that one must be extremely careful in the
physical interpretation of the effective theory.

Let us then  have a closer look at the {\it microscopic} dynamics
that is describes. Effectively, the evolution of the target reduces
to that of a system of {\em dipoles} subjected to a dynamics of a
{\em reaction--diffusion} type: the   dipoles undergo BFKL dynamics,
they can split (one dipole into two dipoles), and they can also
recombine with each other (two dipoles into one). The dynamics of
such a system of dipoles is entirely coded in the $k$--body
densities $n^{(k)}_Y$ (see Sect. 5 in Ref. \cite{IT04} for a precise
definition). Although we shall not work this out explicitly here, it
is not hard, by using the results of Ref. \cite{IM031} to relate
these dipole densities to colorless correlation functions of the
color charge density $\rho^a$. For instance the {\it dipole number
operator} $n(\x,\y)$ can be identified with the bilocal operator
$\rho^a(\x)\rho^a(\y)$ of the effective theory. With such
identifications, and by using Eq.~(\ref{Htotal}), it is
straightforward to construct the evolution equations satisfied by
the dipole densities. One thus finds that $n_Y(\x,\y)$ obeys the
BFKL equation supplemented by a negative term proportional to
$n^{(2)}_Y$, which is generated by the merging piece $H^\dagger_{2
\to 1}$ of the Hamiltonian. Furthermore, the r.h.s. of the equation
for $\del n^{(2)}_Y/\del Y$ includes the standard BFKL terms
describing the individual evolutions of the two dipoles
$(\bm{x}_1,\bm{y}_1)$ and $(\bm{x}_2,\bm{y}_2)$, but also a
positive, {\em fluctuation term}, proportional to $n_Y$ --- this is
generated by the splitting piece $H^\dagger_{1 \to 2}$ of the
Hamiltonian, and is the same as the corresponding term deduced from
the dipole picture in Refs. \cite{IT04,IT05} --- and, finally, a
negative, {\em recombination}, term proportional to $n^{(3)}_Y$. We
thus obtain an infinite hierarchy, which describes a dipole
reaction--diffusion dynamics, as anticipated, and predicts the
saturation of the dipole density at a value of order $1/\alpha_s^2$.

Now, it is clear that this is only an {\em effective dynamics}
since, as well known, dipoles in real QCD do not simply recombine
with each other: the interaction between two dipoles inside the
target wavefunction goes beyond the large--$N_c$ approximation and
leads to more complicated color configurations, involving higher
color multipoles \cite{AM94,AM95}. The reason why it has been
possible to {\em simulate} the non--linear effects responsible for
unitarity corrections in the equations for the scattering
amplitudes through simple `dipole recombination' processes in the
target wavefunction is because the same non--linear effects can be
interpreed as {\em projectile} evolution, in which case they
describe the {\em splitting} of a dipole in the projectile. Then,
the $1 \to 2$ dipole splitting vertex from the projectile is
simply reinterpreted, within the effective theory, as a $2 \to 1$
`dipole merging' vertex in the target. Note finally that a similar
dipole model including splitting and recombination has been
recently used in Ref. \cite{LL05} to generate evolution equations
with Pomeron loops. The present work shows how this {\em
effective} dynamics  may indeed  emerge from the actual target
dynamics in QCD, and points to numerous subtleties involved in
this precise connection.

\vspace*{-0.5cm}
\section*{Acknowledgments}
\vspace*{-0.5cm}

We would like to thank Larry McLerran for sharing with us his intuition
about the potential importance of the fluctuation--saturation duality.
We acknowledge continuous conversations on this and related subjects
with Al Mueller, and useful remarks and comments on the manuscript from
Yoshitaka Hatta and Anna Stasto.

\end{document}